\documentclass[pra,aps,twocolumn,showpacs,longbibliography,floatfix]{revtex4-1}
\usepackage{graphicx}

\newcommand{\ep}{\epsilon}
\newcommand{\vep}{\varepsilon}

\renewcommand{\Im}{\mathop{\mathrm{Im}}\nolimits}
\newcommand{\sign}{\mathop{\mathrm{sign}}\nolimits}
\newcommand{\arcsinh}{\mathop{\mathrm{arcsinh}}\nolimits}
\newcommand{\erf}{\mathop{\mathrm{erf}}\nolimits}
\newcommand{\Ai}{\mathop{\mathrm{Ai}}\nolimits}
\newcommand{\Bi}{\mathop{\mathrm{Bi}}\nolimits}
\newcommand{\md}{\mathrm{d}}

\begin{document}

\title{Landau-Zener-Stueckelberg physics with a singular continuum of states}

\author{D. M. Basko}
\affiliation{
Laboratoire de Physique et Mod\'elisation des Milieux Condens\'es,
Universit\'e Grenoble Alpes and CNRS, F-38000 Grenoble, France}

\begin{abstract}
This work addresses the dynamical quantum problem of a driven discrete
energy level coupled to a semi-infinite continuum whose density of
states has a square-root-type singularity, such as states of a free
particle in one dimension or quasiparticle states in a BCS superconductor.
The system dynamics is strongly affected by the quantum-mechanical
repulsion between the discrete level and the singularity, which gives
rise to a bound state, suppresses the decay into
the continuum, and can produce Stueckelberg oscillations.
This quantum coherence effect may limit the performance of mesoscopic
superconducting devices, such as quantum electron turnstile.
\end{abstract}


\maketitle

Landau-Zener (LZ) transition between two coupled quantum
states whose energies cross in time
is a paradigmatic situation in quantum mechanics.
Due to its generality and simplicity, the LZ model,
originally proposed to describe atomic collisions~%
\cite{Landau1932,Zener1932,Stueckelberg1932}
and spin dynamics in a magnetic field~\cite{Majorana1932},
was later applied to many different phenomena, such as
electron transfer in donor-acceptor complexes~\cite{May2000},
spin dynamics in magnetic molecular clusters~\cite{Wernsdorfer1999},
molecular production in cold atomic gases~\cite{Sun2008},
electron pumping~\cite{Renzoni2001} and
capture~\cite{Kashcheyevs2012} in quantum dots,
dissipation in driven mesoscopic rings~\cite{Gefen1987}
or in superconductor tunnel junctions~\cite{Schon1990,Weissl2015}.
In the course of intense research in various fields, several
generalizations of the two-level LZ model to multiple levels
have been found~%
\cite{Demkov1968,Brundobler1993,Demkov1995,Ostrovsky1997,%
Demkov2001,Pokrovsky2002,Volkov2004,Shytov2004,Patra2015}
including finite-time exact solutions~\cite{Vitanov1996,Mkam2015},
and even many-body versions of the LZ model have been considered~%
\cite{Keeling2008,Altland2008,Sun2008,Ishkhanyan2010}.
However, these generalizations still deal with discrete energy
levels.
A notable exception is Ref.~\cite{Demkov1968}, whose authors
analyzed a single discrete level driven linearly through an
arbitrary spectrum, which could also be continuous. 

\begin{figure}
\includegraphics[width=8cm]{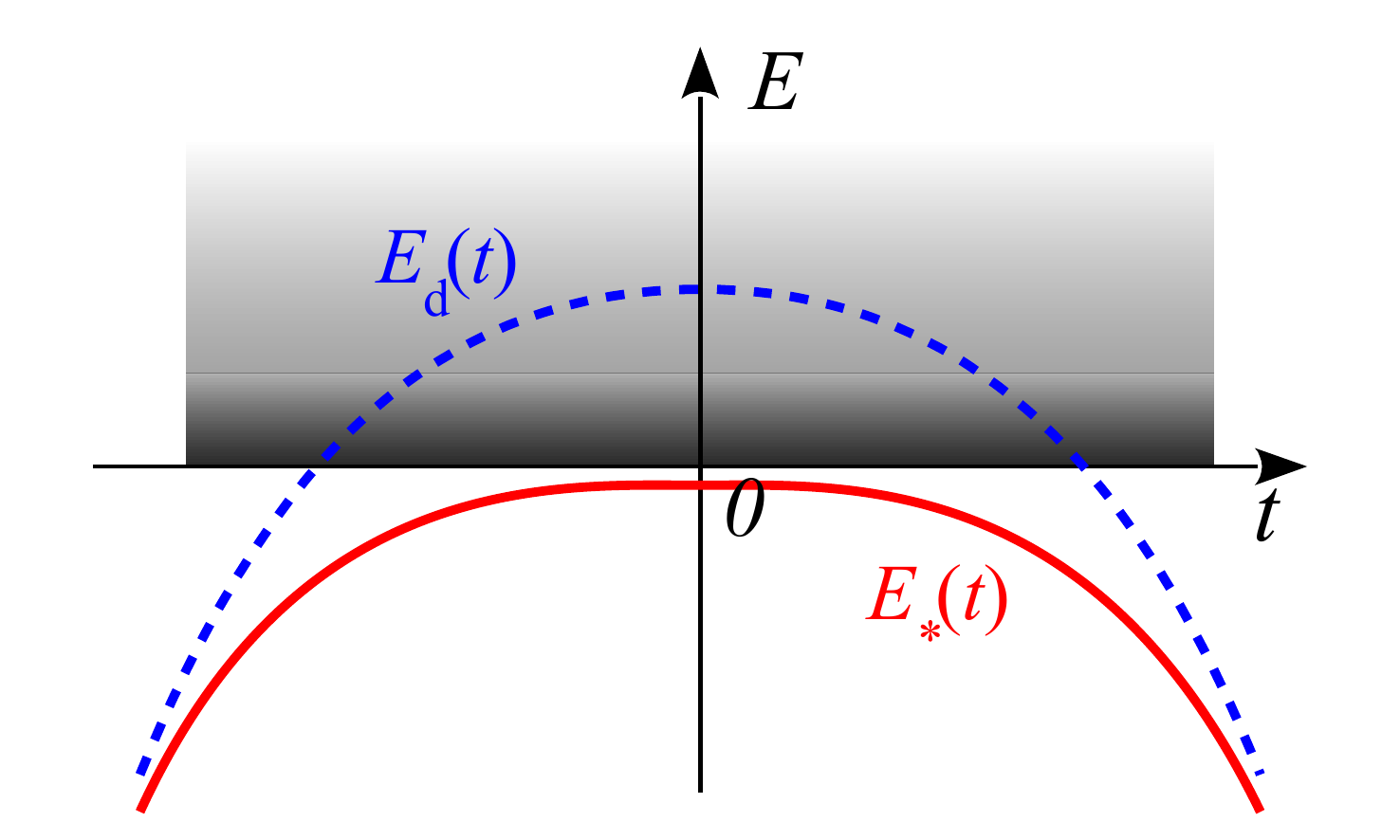}
\caption{\label{fig:levels}(Color online)
A sketch of the time dependence of various energies.
The grey area at $E>0$ represents the continuum with the
singularity in the DOS at $E\to{0}$. The dashed blue line shows the
the bare discrete level $E_\md(t)$, driven inside the continuum
for a finite time. The solid red line shows the adiabatic
ground state $E_*(t)$.
}
\end{figure}

In the present paper, I present another extension of the
Landau-Zener problem involving a discrete level coupled to a
continuum of states, which has an approximate analytical solution
in the long-time limit.
The continuum states are assumed to have positive energies,
$E>0$, with the density of states (DOS) $\nu(E)$ having a
singularity $\nu(E)\propto{1}/\sqrt{E}$ at $E\to{0}^+$.
This singularity is the essential ingredient of the problem.
Physically, such continuum 
can be represented by
a one-dimensional wire with the parabolic dispersion, or
by quasiparticle states in a BCS superconductor above the
superconducting gap. The discrete level (located on an
impurity or a small quantum dot) initially has large
negative energy and contains one particle. Then, its energy
$E_\md$ is moved inside the continuum (e.~g., by applying a
gate voltage), where it stays for some time, and then is
driven back 
to large negative energies,
as shown in Fig.~\ref{fig:levels} by the dashed line.
The quantity of interest is the probability $p_\infty$ for
the particle to stay on the discrete level without being
ejected into the continuum.
A related problem of vanishing bound state in atom-ion
collisions was considered in Ref.~\cite{Demkov1964}.

The practical motivation for the present study comes from the quantum
electron turnstile, a nanoelectronic device transferring
electrons one by one, with a potential metrological
application as a current standard 
(see reviews \cite{Pekola2013,Kaestner2015}).
The electron transfer occurs via a small metallic nanoparticle
sandwiched between two superconducting electrodes~\cite{Pekola2008}.
For a small enough particle, the electron confinement is
very strong, so there is effectively a single electronic level
whose double occupancy is prohibited by the Coulomb repulsion,
and whose energy is controlled by a nearby gate electrode
\cite{vanZanten2015,vanZanten2016}.
The key step of the operation is the electron ejection
from the nanoparticle level, driven by the gate voltage, into the
empty quasiparticle states on one of the superconducting electrodes.
If the superconducting gap is large enough, one can
consider the single-particle problem.
The level trajectory then corresponds to that shown in
Fig.~\ref{fig:levels}, with the energy counted from the BCS
singularity.
The survival probability $p_\infty$
contributes to the turnstile operation error.

The standard description of the decay into a continuum is by
the perturbative Fermi Golden Rule, which gives the decay rate
$\Gamma(E_\md)$ for a fixed level energy~$E_\md$. Application
of the Golden Rule at each instant of time gives
\begin{equation}\label{FGR=}
p_\infty=\exp\left[-\int_{E_\md(t)>0}\Gamma(E_\md(t))\,dt\right],
\end{equation}
where the integration is over the time interval during which
the level stays inside the continuum. Obviously, Eq.~(\ref{FGR=})
is not valid for a too fast drive leading to a large energy
uncertainty. Much less obvious is the breakdown of the
quasistationary Eq.~(\ref{FGR=}) at slow drive. It
is the main focus of the present paper.

The key fact is that for a fixed $E_\md$, the exact eigenstates
of the coupled system form a continuum at $E>0$, and in
addition, there is a discrete bound state at an energy
$E=E_*<0$ \cite{Fetter1965,Machida1972,Shiba1973},
similar to Yu-Shiba-Rusinov states bound to a magnetic
impurity~\cite{Yu1965,Soda1967,Shiba1968,Rusinov1969}.
For large negative~$E_\md$,
the bound state approximately coincides with the
bare discrete level, $E_*\approx{E}_\md$. For $E_\md>0$,
no matter how large, the bound state with $E_*<0$ still
exists, although for $E_\md\to+\infty$ its energy approaches
the continuum and its overlap with the bare discrete state
vanishes.
The existence of the bound state is a consequence of the DOS
singularity at $E\to{0}^+$ and can be viewed as due to the
quantum-mechanical repulsion between the bare level and
the singularity. As the bound state is the adiabatic ground
state of the coupled system, for slow drive the particle will
always stay in it, 
implying $p_\infty\to{1}$.

To describe the crossover between the regime of
Eq.~(\ref{FGR=}) and the adiabatic regime with
$p_\infty\to{1}$, one has to analyze the dynamical problem.
Below, its analytical solution is presented for a special
case of the parabolic time dependence $E_\md(t)$, obtained
by adapting the method of Demkov and Osherov~\cite{Demkov1968}. 
Remarkably, the survival probabilty has the
two-path structure $p_\infty=|A_\md+A_*|^2$, where $A_\md$
corresponds to the resonance in the continuum (the dashed
line in Fig.~\ref{fig:levels}) with $|A_\md|^2$ decaying
according to Eq.~(\ref{FGR=}), while the non-decaying $A_*$
is the contribution of the adiabatic ground state (the solid
line in Fig.~\ref{fig:levels}).
The cross-term in $p_\infty$ describes Stueckelberg-like
interference between the two paths, leading to an oscillatory
dependence of $p_\infty$ on the drive parameters.
Indeed, the bound state may be viewed as a result of avoided
crossing between the discrete level and the singularity; the
double passage of this crossing is similar to Stueckelberg
interferometer.

\emph{The model.}
In a BCS superconductor, the quasiparticle DOS is given by
$\nu(\epsilon)=\nu_0\,\theta(|\ep|-\Delta)|\epsilon|/
\sqrt{\epsilon^2-\Delta^2}$, where $\nu_0$ is the normal-state
DOS, the energy~$\ep$ is counted from the Fermi level,
$2\Delta$ is the superconducting gap, and $\theta(\ep)$~is the
step function.
In the vicinity of the BCS singularity at $\ep\to\Delta$,
the quasiparticle
energy, counted from~$\Delta$ (it is convenient to shift the
energy reference as $E=\ep-\Delta)$, can be approximated as
$E_k=\sqrt{\xi_k^2+\Delta^2}-\Delta\approx\xi_k^2/(2\Delta)$,
and the Bogolyubov quasiparticle factors
$u_k\approx{v}_k\approx{1}/\sqrt{2}$.
Here the index $k$ labels the quasiparticle states, and
$\xi_k$~are the normal-state quasiparticle energies, so that
the state summation $\sum_k$ is represented as $\nu_0\int{d}\xi_k$.
The particle wave
function has a component $\psi_\md$ on the bare discrete level,
and components $\phi_k$ on the continuum states. They satisfy
the two components of the Schr\"odinger equation (we set $\hbar=1$):
\begin{eqnarray}
&&i\,\frac{d\psi_\md}{dt}=E_\md(t)\,\psi_\md
+\sqrt{\frac{\gamma_0}{2\pi\nu_0}}\,\phi_k,\label{ddtpsi=}\\
&&i\,\frac{d\phi_k}{dt}=\frac{\xi_k^2}{2\Delta}\,\phi_k
+\sqrt{\frac{\gamma_0}{2\pi\nu_0}}\,\psi_\md,\label{ddtphi=}
\end{eqnarray}
where the coupling strength is parametrized by~$2\gamma_0$,
the energy-independent decay rate in the normal state.
These equations can be equivalently rewritten
in the coordinate representation, $\phi_k=\int\phi(x)e^{ikx}\,dx$,
$\xi_k^2\to-v_F^2\partial_x^2$, where $v_F$ is the Fermi velocity;
then they become identical to the Schrodinger equation for a simple
one-dimensional wire coupled to a discrete site at $x=0$.

The exact eigenstate energies for fixed~$E_\md$ are found
by substituting $i(d/dt)\to{E}$ and eliminating~$\phi_k$.
This gives the equation $G_\md^{-1}(E)=0$, where the bare
discrete level Green's function and the self-energy are
defined as
\begin{equation}\label{SigmaE=}
G_\md(E)=\frac{1}{E-E_\md-\Sigma(E)},\quad
\Sigma(E)=-\sqrt{\frac{\gamma_0^2\Delta}{-2E}}.
\end{equation}
$\Sigma(E>0)$ is imaginary, describing the particle
escape from the discrete level into the continuum with
the rate $\Gamma(E)=-2\Im\Sigma(E+i0^+)$.
$\Sigma(E<0)$ is real and negative, describing the
quantum-mechanical level repulsion.
The divergence of $\Sigma(E\to{0}^-)$ results in the
existence of a real solution of $G_\md^{-1}(E)=0$ with
$E=E_*<0$ for any~$E_\md$. Thus, the spectrum consists
of a discrete bound state at $E=E_*$, represented by the
isolated pole of $G_\md(E)$, and of the continuum at $E>0$,
corresponding to the branch cut of $\sqrt{-E}$.
The weight of the bare discrete level in the exact bound
state is given by the residue $Z$ of $G_\md(E)$ in the pole
$E=E_*$.
For positive $E_\md\gg(\gamma_0^2\Delta)^{1/3}$, the bound
state is shallow, $E_*\approx-\gamma_0^2\Delta/(2E_\md^2)$,
and the weight is small, $Z\approx\gamma_0^2\Delta/E_\md^3$.

Knowledge of the eigenstates at fixed~$E_\md$ enables
one to treat a special case when the level energy
abruptly rises from $-\infty$ to a finite value~$E_\md$
(a quantum quench), stays constant for a long
time, and then drops back to $-\infty$. The probability
amplitude on the ground state after the first quench is
given by the projection of the discrete level on the
ground state, $\sqrt{Z}$. After a sufficient time the
continuum component is dephased, so on the second quench
the bound state is projected back on the discrete
state, which gives another factor $\sqrt{Z}$. The resulting
survival probability (amplitude squared) is then $p_\infty=Z^2$.

Returning to the dynamical problem, we eliminate
$\phi_k(t)$ from Eqs.~(\ref{ddtpsi=}),  (\ref{ddtphi=}),
and obtain an equation for $\psi_\md(t)$,
\begin{equation}\label{dpsidt=}
i\,\frac{d\psi_\md(t)}{dt}=E_\md(t)\,\psi_\md(t)+
\int_{-\infty}^t\tilde\Sigma(t-t')\,\psi_\md(t')\,dt',
\end{equation}
where
$\tilde\Sigma(t)=e^{5i\pi/4}\theta(t)\sqrt{\gamma_0^2\Delta/(2\pi{t})}$
is the Fourier transform of $\Sigma(E+i0^+)$.
Equation~(\ref{dpsidt=}) should  be solved with the initial
condition $|\psi_\md(t\to-\infty)|=1$, and the quantity of
interest is $p_\infty=|\psi_\md(t\to+\infty)|^2$.

\emph{Markovian regime.}
Let us pass to the interaction representation by writing
$\psi_\md(t)=\Psi_\md(t)\,e^{-i\Phi(t)}$, where 
$\Phi(t)\equiv\int_0^tE_\md(t')\,dt'$.
Equation~(\ref{dpsidt=}) becomes
\begin{equation}\label{RWA=}
i\,\frac{d\Psi_\md(t)}{dt}=
\int_{-\infty}^t\tilde\Sigma(t-t')\,e^{i\Phi(t)-i\Phi(t')}\,\Psi_\md(t')\,dt'.
\end{equation}
If $e^{i\Phi(t)-i\Phi(t')}$ is quickly oscillating for $t'$
far from~$t$, the integral converges at short $t-t'$. If the time
dependence of $\Psi_\md(t')$ is slow enough on the convergence
time scale, one can approximate $\Psi_\md(t')\approx\Psi_\md(t)$
and take it out of the integral (Markovian approximation).
The resulting differential equation is straightforwardly integrated
to give
\begin{equation}\label{pMarkovian=}
p_\infty=\exp\left(-2\int_0^\infty\frac{dE}{2\pi}
\sqrt{\frac{\gamma_0^2\Delta}{2E}}\,|F(E)|^2\right),
\end{equation}
where $F(E)\equiv\int{e}^{iEt-i\Phi(t)}\,dt$.
Equation~(\ref{FGR=}) can be obtained from
Eq.~(\ref{pMarkovian=}) by calculating $F(E)$ in the
stationary phase approximation, or, equivalently, by
approximating $\Phi(t)-\Phi(t')\approx{E}_\md(t)(t-t')$
in Eq.~(\ref{RWA=}), whose right-hand side then becomes
just $\Sigma(E_\md(t))\,\Psi_\md(t)$.

The Markovian character of the integral~(\ref{RWA=})
is lost most easily at times $t\approx{t}_0$ when
$E_\md(t_0)=0$. Approximating
$\Phi(t)\approx\Phi(t_0)+\dot{E}_\md(t_0)\,(t-t_0)^2/2
+\ddot{E}_\md(t_0)\,(t-t_0)^3/3$, where
$\dot{E}_\md\equiv{d}E_\md/dt$, $\ddot{E}_\md\equiv{d}^2E_\md/dt^2$,
we obtain the condition for the validity of the Markovian approximation
as $\max\{|\dot{E}_\md|^{3/2},|\ddot{E}_\md|\}\gg\gamma_0^2\Delta$.
If $E_\md(t)<0$ always, the validity is determined by the values
$E_\md(t_\mathrm{max})$, $\ddot{E}_\md(t_\mathrm{max})$ at the
maximum: $\max\{|E_\md^3|,|\ddot{E}_\md|\}\gg\gamma_0^2\Delta$.

\emph{Adiabatic regime.}
The system is expected to be in the adiabatic regime
as long as $|dE_*/dt|\ll{E}_*^2$ (as in the standard LZ theory).
If this holds at all times, the probability $1-p_\infty$ for the
particle to leave the ground state is expected to be exponentially
small. In this regime, solution of Eq.~(\ref{dpsidt=}), either
analytical or numerical, is not an easy task. Indeed,
Eq.~(\ref{dpsidt=}) is deduced from the Schr\"odinger equation in
the diabatic basis, which is not a natural one to describe the
adiabatic regime~\footnote{The author has not succeeded in obtaining
any result working in the time-dependent adiabatic basis.}. 
Still, by adapting the method of Ref.~\citep{Demkov1968}, an
analytical solution can be found for one specific case of the
parabolic time dependence $E_\md(t)=h-wt^2$, parametrized by
the top energy $h$ and $w>0$ (since $\hbar=1$ was assumed,
$w$~has the dimensionality of energy cubed).

Namely, one goes to the Fourier space,
\begin{equation}\label{Fourier=}
\psi_\md(t)=\int\frac{dE}{2\pi}\,e^{-iEt}\tilde\psi(E),
\end{equation}
where the integration is performed over the real axis.
Since $t^2\to-d^2/dE^2$, Eq.~(\ref{dpsidt=}) is transformed into
\begin{equation}\label{Schroedinger=}
-w\,\frac{d^2\tilde\psi}{dE^2}
+\left[E+\sqrt{\gamma_0^2\Delta/(-2E)}\right]\tilde\psi=h\tilde\psi,
\end{equation}
having the form of the stationary one-dimensional Schr\"odinger
equation with a complex potential (at $E>0$, 
the square root is positive imaginary after analytical continuation
in the upper complex half-plane). The solution must decay
exponentially at $E\to+\infty$. At $E\to-\infty$, it has the
WKB form with some coefficients $C_+,C_-$:
\begin{eqnarray}
&&\tilde\psi(E\to-\infty)=\sum_\pm
C_\pm\frac{e^{\pm{i}S(E)}}{\sqrt{S'(E)}},\quad S'\equiv\frac{dS}{dE},
\label{WKB=}\\
&&S(E)=\int^E\frac{d\vep}{\sqrt{w}}\sqrt{h-\vep-\sqrt{\gamma_0^2\Delta/(-2\vep)}}.
\end{eqnarray}
At $t\to\pm\infty$, the integral in Eq.~(\ref{Fourier=}) can be calculated
in the stationary phase approximation. For each~$t$, only one of the
two terms in Eq.~(\ref{WKB=}) produces a stationary point, determined
by $\pm{t}=S'(E)>0$. At $|t|\to\infty$, the solution
$E_t=h-wt^2\to-\infty$, so one can indeed use the asymptotic WKB
expression~(\ref{WKB=}). As a result,
\begin{equation}
\psi_\md(t\to\pm\infty)=e^{\mp{i}\pi/4}\sqrt{\frac{w}\pi}\,
C_\pm{e}^{-iE_tt\pm{i}S(E_t)},
\end{equation}
which gives $p_\infty=|C_+/C_-|^2$. Thus, the survival probability
$p_\infty$ of the dynamical problem~(\ref{dpsidt=}) corresponds
to the inverse reflection coefficient in the scattering problem
for the Schr\"odinger equation~(\ref{Schroedinger=}).
The positive imaginary part of the potential ensures $p_\infty<1$.

\begin{figure}
\includegraphics[width=8cm]{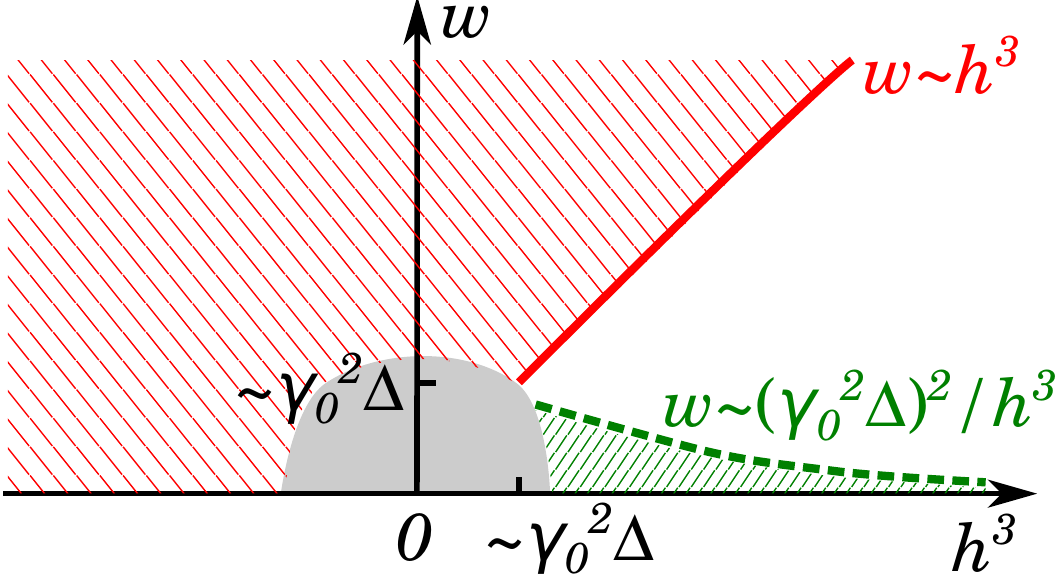}
\caption{\label{fig:regimes} (Color online)
Different regimes for the problem~(\ref{dpsidt=})
with $E_\md(t)=h-wt^2$.
The adibatic regime with $1-p_\infty\ll{1}$ (hatched area
below the dashed line) occurs if the condition
$|dE_*/dt|\ll{E}_*^2$ holds at all times.
In the fast-drive regime (hatched area to the left of the
solid line), the time spent in the continuum is too short,
so that $h$ is within the energy uncertainty
and $1-p_\infty\ll{1}$.
In the Golden-Rule regime
(white area between the dashed and the solid line) $p_\infty\ll{1}$.
The grey area corresponds to $h^3\sim{w}\sim\gamma_0^2\Delta$
with $p_\infty\sim{1}$.
}
\end{figure}

The adiabatic effect is nontrivial when the bound state is shallow,
$h\gg(\gamma_0^2\Delta)^{1/3}$.
We also assume $h^3\gg{w}$; otherwise, the time spent by
$E_\md(t)$ in the continuum is too short (the energy uncertainty
exceeds~$h$), and $p_\infty\approx{1}$
can be found from Eq.~(\ref{pMarkovian=}) with
$F(E)=2\pi{w}^{-1/3}\Ai({w}^{-1/3}(E-h))$, where $\Ai(x)$ is the
Airy function.
When $h^3\gg{w},\gamma_0^2\Delta$ (below the red solid line
in Fig.~\ref{fig:regimes}), the wave function $\tilde\psi(E)$
can be found in the WKB approximation everywhere except (i)~the
vicinity of the classical turning point $E=h$, where it can be
treated in the standard way, and (ii)~near the singularity at
$E\to{0}$ (see Supplemental Material for details).
Then, one can identify two limiting cases for matching the
WKB solution at $E\to{0}$, governed by the parameter
$\sqrt{wh^3}/(\gamma_0^2\Delta)$. They are separated by the dashed
line in Fig.~\ref{fig:regimes}.

(i) In the adiabatic regime, $\sqrt{wh^3}\gg\gamma_0^2\Delta$, the
particle stays in the ground state up to an exponentially small
ejection probability,
\begin{equation}\label{padiabatic=}
p_\infty=1-\exp
\left(-\frac\pi{4}\,\frac{\gamma_0^2\Delta}{\sqrt{wh^3}}\right).
\end{equation}

(ii) In the opposite limit, $C_+/C_-$ is calculated to the first order
in $\sqrt{wh^3}/(\gamma_0^2\Delta)\ll{1}$, which gives
\begin{equation}\label{pcrossover=}
p_\infty=\left|e^{-\pi\sqrt{\gamma_0^2\Delta/(2w)}-(4/3)i\sqrt{h^3/w}}
+e^{i\pi/4}\sqrt{\frac\pi{4}\,\frac{\gamma_0^2\Delta}{\sqrt{wh^3}}}\right|^2.
\end{equation}
The first term [of zero order in 
$\sqrt{wh^3}/(\gamma_0^2\Delta)$] gives the Golden-Rule
expression~(\ref{FGR=}); indeed, the exponent is nothing but
$(1/2)\int\Gamma(E_\md(t))\,dt-i\int{E}_\md(t)\,dt$ for $E_\md(t)=h-wt^2$.
The second term is the first-order correction which must be small
compared to unity, but can still be larger than the zero-order term.
Remarkably, in the latter case it matches the adiabatic
expression~(\ref{padiabatic=}) obtained in the opposite limit.

\emph{Discussion.}
Equations (\ref{pMarkovian=}), (\ref{padiabatic=}) and (\ref{pcrossover=})
represent the main result of the present work.
They agree with the numerical solution of Eq.~(\ref{dpsidt=}) (see the Supplemental Material). 
Although Eqs.~(\ref{padiabatic=}) and (\ref{pcrossover=}) are obtained
for a specific dependence $E_\md(t)=h-wt^2$, their relevance is quite general,
since any smooth $E_\md(t)$ can be approximated by a parabola near the maximum.
The three expressions have overlapping domains of validity:
Eq.~(\ref{pMarkovian=}) with $F(E)=2\pi{w}^{-1/3}\Ai({w}^{-1/3}(E-h))$
matches the first term in Eq.~(\ref{pcrossover=}), while
Eq.~(\ref{padiabatic=}) matches the second.
The only region not covered by
Eqs.~(\ref{pMarkovian=}), (\ref{padiabatic=}) and (\ref{pcrossover=})
is $h^3\sim{w}\sim\gamma_0^2\Delta$, shown in
Fig.~\ref{fig:regimes} by the grey area.

Eq.~(\ref{pcrossover=}) has a two-path form, corresponding to the
two trajectories shown in Fig.~\ref{fig:levels}.
Due to the $h$- and $w$-dependent phase of the first term,
$p_\infty$~may exhibit Stueckelberg interference oscillations as a
function of~$h$ or~$w$.
From the analogy with the standard two-level problem, it is tempting
to assume that the crossing of the singularity at $t_0=-\sqrt{h/w}$
can be viewed as a beam splitter, when the particle ``decides'' which
path to follow.
However, if this were the case, the system behavior would be
determined by $E_\md(t)$ linearized around~$t_0$, i.~e.,
by $\dot{E}_\md(t_0)=2\sqrt{wh}$, while in Eq.~(\ref{pcrossover=})
the parameter governing the amplitude of the adiabatic path is
$\sqrt{wh^3}/(\gamma_0^2\Delta)$.
This latter parameter is nothing but the maximal value of
$E_*^{-2}|dE_*/dt|$, which should be small to keep the
adiabaticity at all times. This maximal value is reached at
$t\approx t_0/\sqrt{3}$, quite far from the crossing.

In any realistic superconduncting device, the BCS singularity
in the DOS, which is the key ingredient of the problem,
is necessarily smeared on some energy scale. If the smearing
exceeds $|E_*|$, the bound state enters the continuum and
decays, so the described effect is no longer relevant.
The smearing is often quantified by the Dynes
parameter~\cite{Dynes1978,Dynes1984} which gives the ratio of
the smearing scale to the gap~$\Delta$.
For aluminum-based superconducting nanostructures, the Dynes
parameter is typically $10^{-4}-10^{-5}$, mostly due to
microwave noise from the environment~\cite{Pekola2010}, and
can be made as low as $10^{-7}$ if special efforts are made
to ensure efficient microwave shielding and quasiparticle
relaxation~\cite{Saira2012}.
Taking the values $\gamma_0=1\:\mu\mbox{eV}$,
$\Delta=200\:\mu\mbox{eV}$~\cite{vanZanten2015}, 
we obtain the main energy scale
responsible for the formation of the bound state
$(\gamma_0^2\Delta/2)^{1/3}\approx{5}\:\mu\mbox{eV}$,
which exceeds the Dynes smearing by several orders of magnitude.
For a sinusoidal drive with the amplitude $100\:\mu\mbox{eV}$
and frequency $50\:\mbox{MHz}$~\cite{vanZanten2016}, we obtain
$w\approx{2}\:\mu\mbox{eV}^3$.
Then the level should be pushed by
$h\sim[(\gamma_0^2\Delta/2)^2/w]^{1/3}\sim$ a few $\mu${eV}
beyond the BCS singularity to overcome the adiabatic blocking,
and the period of the Stueckelberg oscillations is
$h\sim{w}^{1/3}\sim{1}\:\mu\mbox{eV}$, both corresponding to
quite measurable energy scales. To give a noticeable amplitude
of the oscillations, $w$~should not be too small compared to
$\pi^2\gamma_0^2\Delta/2$, so it is better to use a device with
sub-$\mu\mbox{eV}$~$\gamma_0$.

The experimental resolution is more likely to be limited by
the high-frequency noise component of the driven gate voltage,
which should favor electron ejection from the bound state into
the continuum. Thus, in experiment, special care should be
taken in order to reduce this extrinsic noise. Theoretically,
the effect of noise has been studied for the standard two-level
Landau-Zener problem
\cite{Shimshoni1991,Shytov2003,Pokrovsky2003,Vestgarden2008,Whitney2011,Kenmoe2013};
inclusion of noise in the present theory along the same lines
is a subject for the future work.

To conclude, I presented an extension of the Landau-Zener problem
to a continuous energy spectrum.
The key role is played by the singularity
in the continuum DOS, which is crossed by the driven discrete
level. The Landau-Zener physics is not washed out
by the continuum because of the quantum-mechanical level
repulsion between the discrete level and the DOS singularity,
and even Stueckelberg oscillations are present.
The fundamental physics, described here, is shown to be
relevant for a specific mesoscopic device, the hybrid quantum
electron turnstile, where the BCS singularity in the
quasiparticle DOS of superconducting electrodes may prevent
electron ejection from the discrete quantum dot level into
the electrode, thereby providing a fundamental limit on the
device operation.

\section*{Acknowledgements}
The author is grateful to D.~Van Zanten, C.~Winkelmann, and H.~Courtois for the stimulating discussions which initiated this work, as well as to Yu.~Galperin, M.~Houzet, I.~Khaymovich, 
M.~Kiselev, L.~Levitov, J.~Pekola, V.~Pokrovsky, A.~Shushin, X.~Waintal, 
R.~Whitney, E.~Yuzbashyan, and many others
for helpful discussions on various stages of the work.

\bibliography{LandauZener}

\clearpage

\begin{widetext}
\begin{center}
{\large\bf Supplemental Material}
\end{center}
\end{widetext}

\section*{Analytical solution of the stationary Schr\"odinger equation}
\label{app:analytical}

\begin{figure}
\includegraphics[width=7cm]{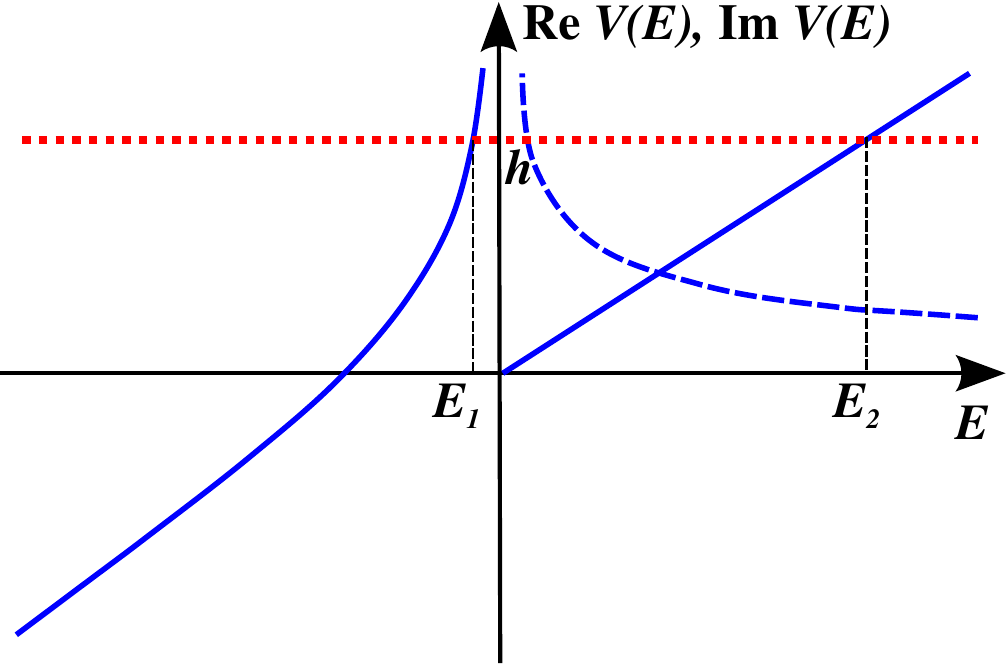}
\caption{\label{fig:potenital} (Color online)
The real and imaginary parts of the potential $V(E)$
(the blue solid and dashed curves, respectively).
The energy $h$ is shown by the horizontal red dotted line.
The scattering problem corresponds to the wave incident from
the left, tunnelling under the square-root spike, and
propagating until $E=E_2$, where it is necessarily reflected.
While propagating at $E>0$, the wave is amplified due to
$\Im{V}(E)>0$.
}
\end{figure}

Here we study the stationary Schr\"odinger equation,
\begin{equation}
-w\,\frac{d^2\tilde\psi(E)}{dE^2}
+V(E)\,\tilde\psi(E)=h\tilde\psi(E),
\end{equation}
where $E$ plays the role of the coordinate,
$1/(2w)$ and $h$ represent the mass and the energy,
respectively, and
\begin{equation}\label{potential=}
V(E)=E+\sqrt{\frac{\gamma_0^2\Delta}{-2E}}
\end{equation}
is the effective potential, plotted in Fig.~\ref{fig:potenital}.
$V(E)$ is real at $E<0$, while at $E>0$ the square root should
be analytically continued in the upper complex half-plane, giving
$\Im{V}(E)>0$. The probability current, defined as
\begin{equation}
J(E)=-iw\left[\tilde\psi^*(E)\,\frac{d\tilde\psi(E)}{dE}
-\frac{d\tilde\psi^*(E)}{dE}\,\tilde\psi(E)\right],
\end{equation}
satisfies the continuity equation,
\begin{equation}\label{continuity=}
\frac{dJ(E)}{dE}=2\Im{V}(E)\,|\tilde\psi(E)|^2.
\end{equation}
Since $\tilde\psi(E\to+\infty)$ must be exponentially decaying,
Eq.~(\ref{continuity=}) implies $J(E)<0$, which ensures
$p_\infty<1$.

In the following, we assume $h^3\gg{w},\gamma_0^2\Delta$.
Then on most of the real axis, the solution can be approximated
by the WKB form:
\begin{eqnarray}
&&\tilde\psi(E\to-\infty)=\left(\frac{h}w\right)^{1/4}\sum_\pm
C_\pm\frac{e^{\pm{i}S(E)}}{\sqrt{S'(E)}},\label{tpsiWKB=}\\
&&S(E)=\int_{E_1}^E\sqrt{\frac{h-V(\varepsilon)}w}\,d\varepsilon,
\label{SE=}
\end{eqnarray}
where $S'(E)\equiv{dS}/dE$, and we choose the lower limit $E_1$ to
be the leftmost classical turning point of $V(E)$, where
$V(E)=h$. Potential~(\ref{potential=}) has two turning points:
\begin{equation}
E_1=-\frac{\gamma_0^2\Delta}{2h^2}+O((\gamma_0^2\Delta)/h^5),
\quad
E_2=h+O\!\left(\sqrt{\gamma_0^2\Delta/h}\right)
\end{equation}
(note that $E_2$ is complex).
The conditions $h^3\gg{w},\gamma_0^2\Delta$ ensure that $E_1$
and $E_2$ are well separated, so that one can use the WKB
expression in the region $0<E<E_2$.
In fact, $E_1$ and $E_2$ are nothing but the poles of the Green's
function $G_\md(E)$ from Eq.~(\ref{SigmaE=}) for $E_\md=h$, so the
considered regime correponds to well separated peaks in the
discrete level spectral function.
 
Let us introduce two pairs of WKB solutions, $\xi_\pm(E)$ and
$\eta_\pm(E)$, representing the right/left-traveling waves
in the regions $E<E_1$ and $0<E<E_2$, respectively:
\begin{eqnarray}
&&\xi_\pm(E<E_1)=\left(\frac{h}w\right)^{1/4}
\frac{e^{\pm{i}S(E)}}{\sqrt{S'(E)}},\\
&&\eta_\pm(0<E<E_2)=\left(\frac{h}w\right)^{1/4}
\frac{e^{\pm{i}S(E)}}{\sqrt{S'(E)}}.
\end{eqnarray}
While $S(E<E_1)$ is defined by the integral in Eq.~(\ref{SE=})
on the real axis, $S(0<E<E_2)$ should be understood as the
analytical continuation from $E<E_1$ through the upper complex
half-plane. The two pairs of solutions must be linear combinations
of each other, so we can define a transfer matrix
$\mathcal{T}_{\kappa\kappa'}$ with $\kappa,\kappa'=\pm$, such that
\begin{equation}
\eta_\kappa(E)=\sum_{\kappa'=\pm}
\mathcal{T}_{\kappa\kappa'}\xi_{\kappa'}(E).
\end{equation}
The Wronskian conservation,
$W(\xi_+,\xi_-)=W(\eta_+,\eta_-)=-2i\sqrt{h/w}$,
imposes $\det\mathcal{T}=1$.

At the turning point $E=E_2$ (even for complex~$E_2$), $V(E)$
can be linearized, so the solution decaying at $E\to+\infty$
is constructed by the standard WKB prescription:
\begin{eqnarray}
&&\tilde\psi(E)=\frac{2(h/w)^{1/4}}{\sqrt{S'(E)}}
\cos\left[S(E)-S(E_2)+\frac\pi{4}\right]
=\nonumber\\
&&\qquad{}=\sum_{\kappa=\pm}e^{i\kappa[\pi/4-S(E_2)]}\eta_\kappa(E)
=\nonumber\\
&&\qquad{}=\sum_{\kappa,\kappa'=\pm}e^{i\kappa[\pi/4-S(E_2)]}
\mathcal{T}_{\kappa\kappa'}\xi_{\kappa'}(E).
\label{tpsi=xixi}
\end{eqnarray}
As discussed in the main text, $p_\infty$ is determined by the
ratio of the coefficients at $\xi_+(E)$ and $\xi_-(E)$:
\begin{equation}\label{pinfty=}
p_\infty=\left|\frac{\mathcal{T}_{-+}+i\mathcal{T}_{++}e^{-2iS(E_2)}}%
{\mathcal{T}_{--}+i\mathcal{T}_{+-}e^{-2iS(E_2)}}\right|^2.
\end{equation}
This reduces the problem to (i)~evaluating $S(E_2)$ and
(ii)~finding the matrix $\mathcal{T}_{\kappa\kappa'}$, determined
by the scattering on the singularity of the potential~$V(E)$.

To find $S(E_2)$, let us expand
\begin{equation}
S'(E)=\sqrt{\frac{h-E}w-\sqrt{\frac{\gamma_0^2\Delta}{-2w^2E}}}
\end{equation}
in $\sqrt{\gamma_0^2\Delta}$, and integrate it term by term:
\begin{eqnarray}
&&S(E)=\frac{2}{3}\sqrt{\frac{h^3}{w}}
-\frac{2}{3}\sqrt{\frac{(h-E)^3}{w}}
+{}\nonumber\\ &&\quad\qquad
{}+\sqrt{\frac{\gamma_0^2\Delta}{2w}}\arcsinh\sqrt{-\frac{E}{h}}
+{}\nonumber\\ &&\quad\qquad
{}+O\!\left(\sqrt{\frac{\gamma_0^2\Delta}{w}}
\sqrt{\frac{\gamma_0^2\Delta}{(h-E)^3}}\,\ln(-E)\right).
\label{SE1=}
\end{eqnarray}
This expression describes $S(E)$ at $E<E_1$, while for $E>0$ the
analytical continuation gives $\sqrt{-E}\to-i\sqrt{E}$,
$\arcsinh\sqrt{-E/h}\to-i\arcsin\sqrt{E/h}$. This imaginary term
in $S(E)$ produces the enhancement of the propagating waves in
the region $0<E<h$, required by Eq.~(\ref{continuity=}).
Expansion (\ref{SE1=}) breaks down at $E\to{h}$. Let us now
expand around $E=E_2$:
\begin{eqnarray}
&&S(E)=S(E_2)-\frac{2}{3}
\left(1-\frac{i}4\sqrt{\frac{\gamma_0^2\Delta}{2h^3}}\right)
\sqrt{\frac{(E_2-E)^3}{w}}
+{}\nonumber\\ &&\quad\qquad
{}+O\!\left(\sqrt{\frac{\gamma_0^2\Delta}{w}}
\sqrt{\frac{(E_2-E)^5}{h^5}}\right).\label{SE2=}
\end{eqnarray}
Expansion~(\ref{SE1=}) assumes $h-E\gg\sqrt{\gamma_0^2\Delta/h}$,
while expansion~(\ref{SE2=}) assumes $|E_2-E|\ll{h}$, so they can
be matched 
in the parametrically wide region where both inequalities are
satisfied. This gives the leading real and imaginary terms
in $S(E_2)$:
\begin{equation}
S(E_2)\approx\frac{2}{3}\sqrt{\frac{h^3}w}
-\frac{i\pi}2\sqrt{\frac{\gamma_0^2\Delta}{2w}}.
\end{equation}
Note that subleading terms
$\sim\gamma_0^2\Delta/\sqrt{wh^3}$ can still be larger than
unity. It will be seen below that the precise value of $S(E_2)$
is not important for $wh^3\ll(\gamma_0^2\Delta)^2$.

To determine the matrix $\mathcal{T}_{\kappa\kappa'}$, one can
neglect the linear term $E$ in the potential $V(E)$ because
$|E_1|\ll{h}$. Then, it is convenient to rescale the energy,
$E=y\sqrt{w/h}$, and rewrite the Schr\"odinger equation as
\begin{equation}\label{sqrtSchr=}
-\frac{d^2\tilde\psi}{dy^2}+\frac{\alpha}{\sqrt{-y}}\,\tilde\psi
=\tilde\psi,
\quad\alpha\equiv\sqrt{\frac{\gamma_0^2\Delta/2}{\sqrt{wh^3}}}.
\end{equation}
This equation has an exact solution, expressed in
terms of the confluent hypergeometric function~\cite{Ishkhanyan2015}.
Still, the two limiting cases $\alpha\ll{1}$ and $\alpha\gg{1}$ can
be analyzed without invoking the exact solution. This is done below,
and simple expressions for $p_\infty$ are obtained.

For $\alpha=0$ one trivially obtains 
$\mathcal{T}_{\kappa\kappa'}=\delta_{\kappa\kappa'}$.
When substituted in Eq.~(\ref{pinfty=}), it gives the first term
in Eq.~(\ref{pcrossover=}), that is, the Golden-Rule result~(\ref{FGR=}).

For $\alpha\ll{1}$ one can calculate the first perturbative correction
to $\mathcal{T}_{\kappa\kappa'}=\delta_{\kappa\kappa'}$. Let us look
for two linearly independent solutions of Eq.~(\ref{sqrtSchr=}) in the
form $\tilde\psi(y)=[1+\alpha{u}(y)]e^{\pm{iy}}$. Then $u(y)$
satisfies $u''\pm{2}iu'=1/\sqrt{-y}$, and writing further
$u'(y)=v(y)\,e^{\mp{2}iy}$, we obtain the wave functions in the form
\begin{equation}\label{y1y2integral=}
\tilde\psi(y)=e^{\pm{i}y}+\alpha{}e^{\pm{i}y}\int\limits^ydy_1
\int\limits^{y_1}dy_2\,
\frac{e^{\pm{2}i(y_2-y_1)}}{\sqrt{-y_2}}+O(\alpha^2).
\end{equation}
Taking different lower integration limits corresponds to forming
different linear combinations of $\xi_\pm(y)$ or $\eta_\pm(y)$,
and one is free to choose the most convenient one. Indeed, to find
the matrix $\mathcal{T}_{\kappa\kappa'}$ it is sufficient to
construct any pair of linearly independent solutions and to match
it to $\xi_\pm$ and $\eta_\pm$. Choosing both lower limits to be
zero and integrating over $y_1$ by parts, one readily obtains
a compact expression in terms of the error function:
\begin{eqnarray}
&&\tilde\psi(y)=e^{\pm{i}y}\pm{i}\alpha\sqrt{-y}\,{}e^{\pm{i}y}
+{}\nonumber\\ &&\qquad\quad{}+
\alpha\sqrt{\frac\pi{8}}\,e^{\mp{3}i\pi/4\mp{i}y}
\erf\!\left(\sqrt{-2y}\,e^{\pm{i}\pi/4}\right)+O(\alpha^2).\nonumber\\
\label{tpsiexpansion=}
\end{eqnarray}
Let us now write the expansion of the WKB solutions
$(1-\alpha/\sqrt{-y})^{-1/4}e^{\pm{i}S(y)}$ with
$S(y)=y+\alpha\sqrt{-y}$~(\ref{SE1=}):
\begin{eqnarray}
&&\xi_\pm(y<0)=\left(1+\frac{1}{4}\,\frac{\alpha}{\sqrt{-y}}
\pm{i}\alpha\sqrt{-y}\right)e^{\pm{i}y}+O(\alpha^2),\nonumber\\
&&\label{xialphaexp=}\\
&&\eta_\pm(y>0)=\left(1+\frac{i}{4}\,\frac{\alpha}{\sqrt{y}}
\pm\alpha\sqrt{y}\right)e^{\pm{i}y}+O(\alpha^2).\label{etaalphaexp=}
\end{eqnarray}
To match them to Eq.~(\ref{tpsiexpansion=}), one
should use the asymptotic expression for $\erf(z)$ paying
attention to the essential singularity at $z=\infty$,
so that for real $s\to+\infty$
\[
\erf(se^{\pm{i}\pi/4})=1-\frac{e^{\mp{is^2}\mp{i}\pi/4}}{\sqrt\pi{s}}+O(1/s^2),
\]
but at the same time, due to $\erf(-z)=-\erf(z)$,
\[
\erf(se^{-3i\pi/4})=
-1+\frac{e^{-is^2-i\pi/4}}{\sqrt\pi{s}}+O(1/s^2).
\]
The result is
\begin{eqnarray}
&&\tilde\psi(y)
=\xi_\pm(y)+\alpha\sqrt{\frac\pi{8}}\,e^{\mp{3}i\pi/4}\xi_\mp(y)+O(\alpha^2)
=\nonumber\\
&&\qquad{}=
\eta_\pm(y)\pm\alpha\sqrt{\frac\pi{8}}\,e^{\mp{3}i\pi/4}\eta_\mp(y)+O(\alpha^2),
\end{eqnarray}
from which 
$\mathcal{T}_{\kappa\kappa'}$ is obtained
to the first order in~$\alpha$:
\begin{eqnarray}
&&\mathcal{T}_{++},\mathcal{T}_{--}=1+O(\alpha^2),\quad
\mathcal{T}_{+-}=O(\alpha^2),\\
&&\mathcal{T}_{-+}=\alpha\sqrt{\frac\pi{2}}\,e^{3i\pi/4}+O(\alpha^2).
\end{eqnarray}
Its substitution into~(\ref{pinfty=}) gives the second term
in~(\ref{pcrossover=}).

For $\alpha\gg{1}$, the key observation is that the classical turning
point $y=-\alpha^2$ lies quite far from the singularity at $y=0$,
so there is a wide classically forbidden region between $-\alpha^2$
and~0. As a result, only an exponentially small part of the incident
wave will be able to tunnel to the amplification zone at $y>0$.
Moreover, in most of the classically forbidden region the WKB
approximation can be used, so up to exponentially small terms the
sought solution~$\tilde\psi(y)$ can be written in the standard
WKB form:
\begin{eqnarray}
&&\tilde\psi(y<-\alpha^2)=e^{i\pi/4}\xi_+(y)+e^{-i\pi/4}\xi_-(y),\\
&&\tilde\psi(-\alpha^2<y<0)=\frac{e^{-\sigma(y)}}{\sqrt{\sigma'(y)}},
\label{tpsi=esigma}
\end{eqnarray}
where $\sigma'(y)=d\sigma(y)/dy$ and $\sigma(y)$ is just $\Im{S}$
in the forbidden region:
\begin{eqnarray}
&&\sigma(y)=\int\limits_{-\alpha^2}^y\sqrt{\frac\alpha{\sqrt{-y_1}}-1}\,dy_1
={}\nonumber\\
&&\qquad{}=\frac{\alpha^2}4\arccos\left(\frac{2\sqrt{-y}}\alpha-1\right)
-{}\nonumber\\
&&\qquad\quad{}-\left(\sqrt{-y}-\frac\alpha{2}\right)\sqrt{\alpha\sqrt{-y}+y}.
\end{eqnarray}
Keeping the exponentially growing solution $\propto{e}^{\sigma(y)}$
in the whole region $-a^2<y<0$ is beyond the WKB accuracy. However,
one should keep in mind that at $y\to{0}$ its amplitude is of the
same order as that of solution~(\ref{tpsi=esigma}).
At $|y|\sim\alpha^{-2/3}$,
the WKB approximation breaks down; however, at $|y|\ll\alpha^2$ one
can neglect the right-hand side of Eq.~(\ref{sqrtSchr=}) and solve it
exactly, obtaining two linearly independent solutions.
At positive $y\gg\alpha^{-2/3}$, the WKB
approximation is again valid, and
the sought solution $\tilde\psi(y)$ is a linear combination of
$\eta_\pm(y)$, with the amplitude of $\eta_+(y)$ being exponentially
smaller than that of $\eta_-(y)$, according to Eq.~(\ref{tpsi=xixi}).
Thus, one can neglect the $\eta_+$ component and solve
Eq.~(\ref{sqrtSchr=}) with zero right-hand side and with the boundary
condition of exponentially decaying $\tilde\psi(y)\propto\eta_-(y)$
at $y\to+\infty$. At $y\to-\infty$ the solution will have both
$e^{\sigma(y)}$ and $e^{-\sigma(y)}$ components, and the amplitude
of the latter is given by Eq.~(\ref{tpsi=esigma}).

\begin{figure}[t]
\includegraphics[width=7cm]{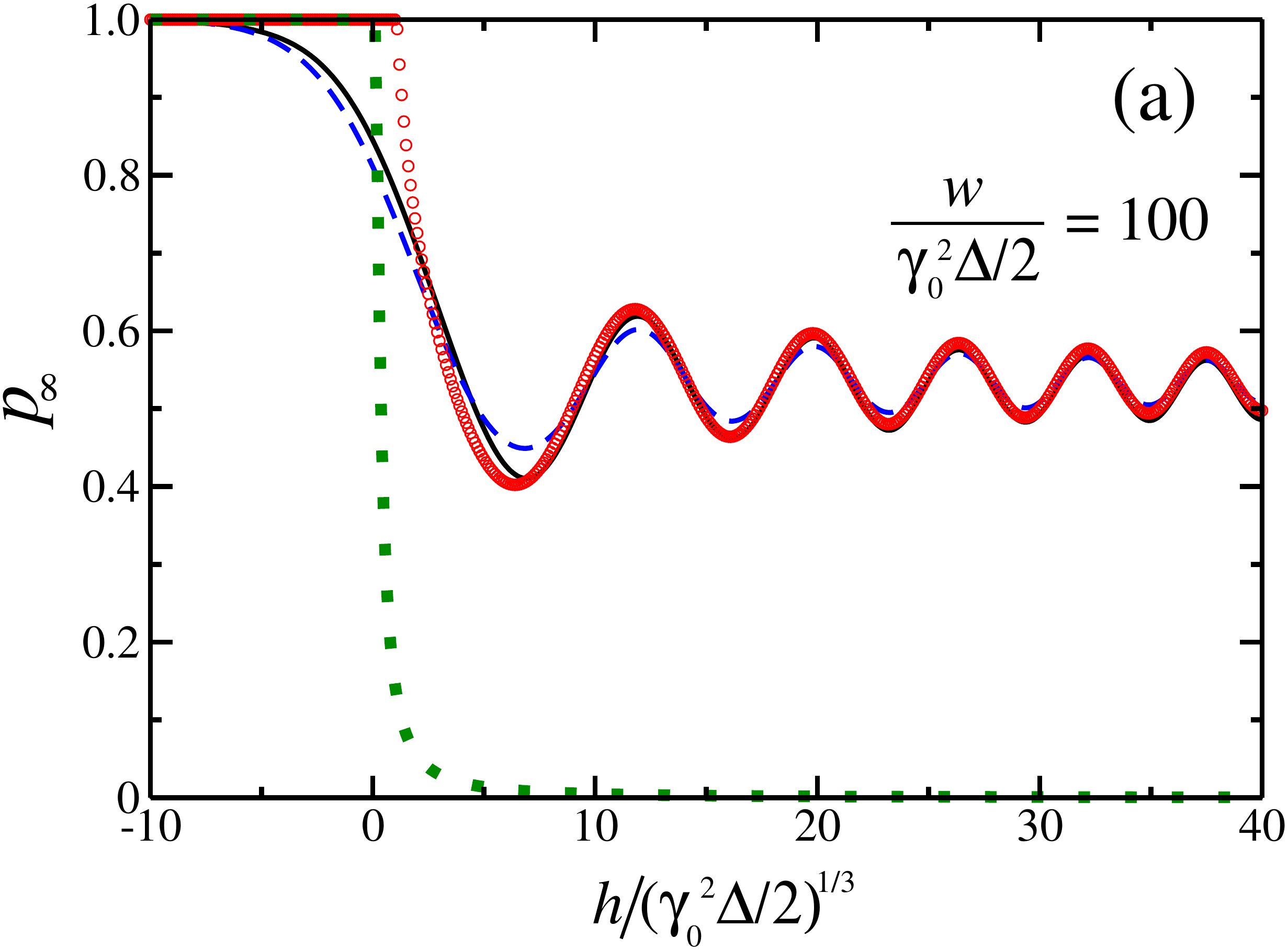}\vspace{0.5cm}\\
\includegraphics[width=7cm]{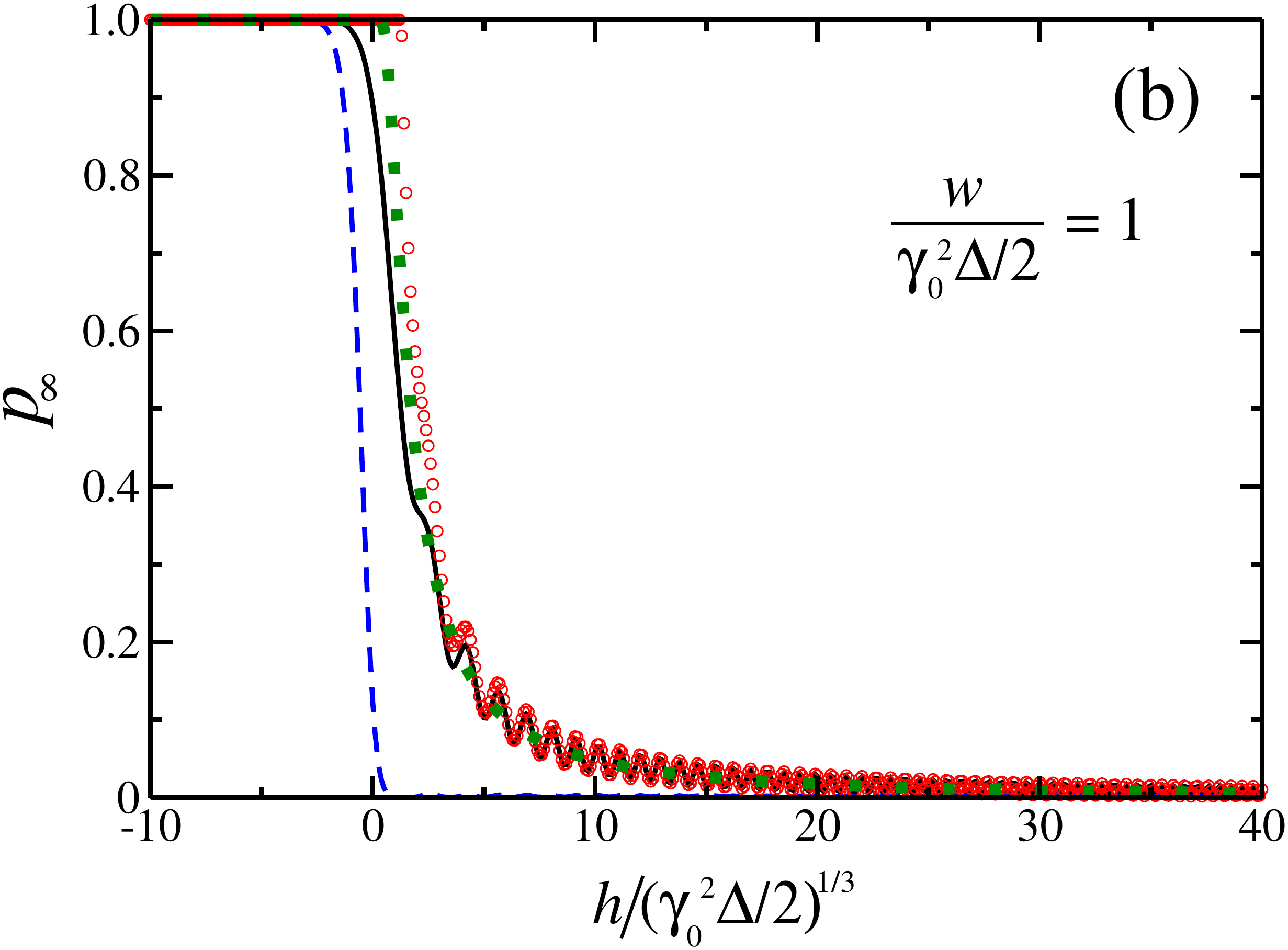}\vspace{0.5cm}\\
\includegraphics[width=7cm]{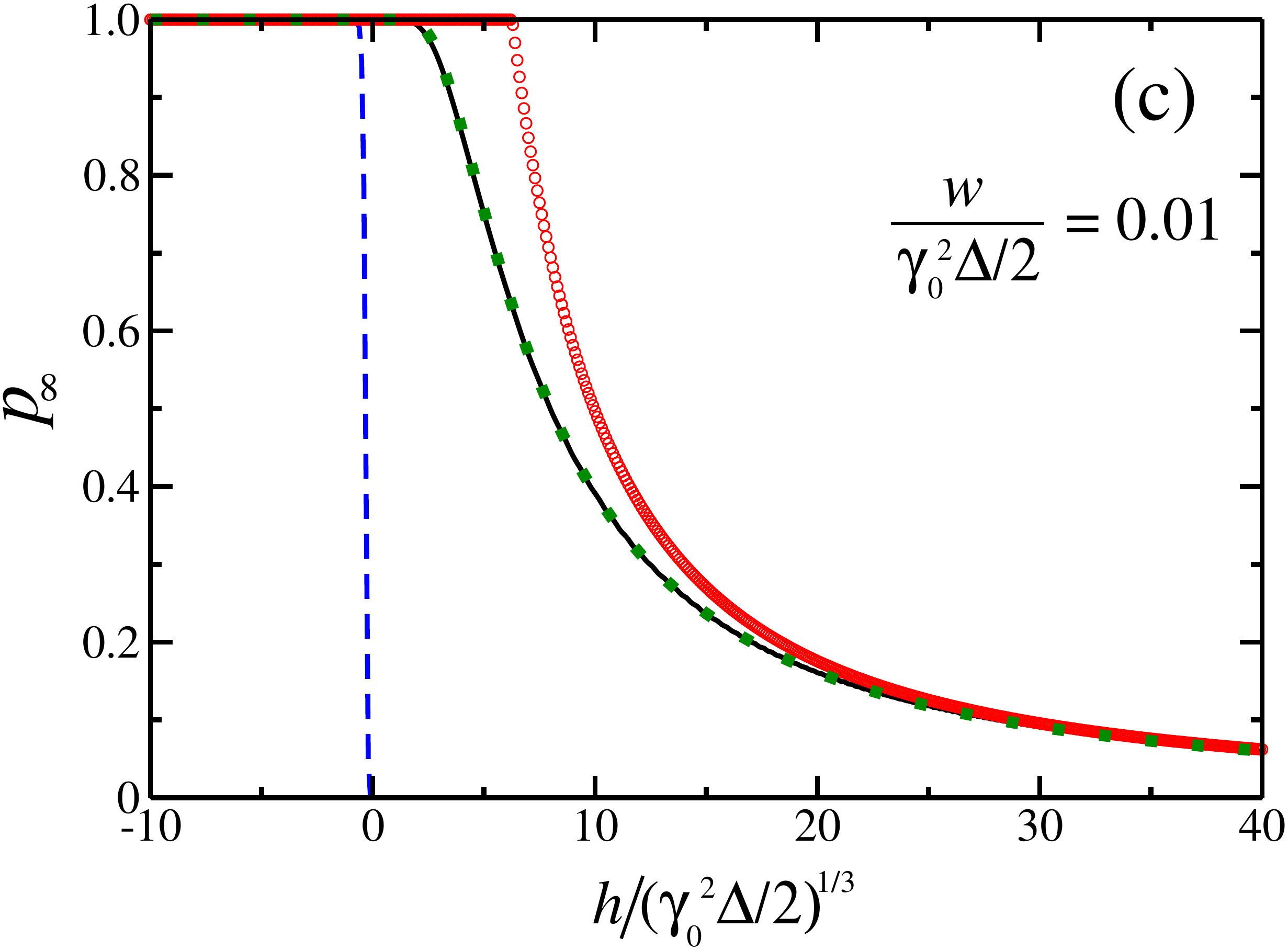}
\caption{\label{fig:curves} (Color online)
Comparison between the different analytical expressions for $p_\infty$
from the main text and the results of the numerical integration
for $w/(\gamma_0^2\Delta/2)=100,\,1,\,0.01$, shown on panels
(a), (b), (c), respectively.
The black solid line represents the numerical result.
The dashed blue line is the Markovian result, Eq.~(\ref{pMarkovian=}).
The green squares represent the adiabatic result,
Eq.~(\ref{padiabatic=}).
The red circles show the crossover expression, Eq.~(\ref{pcrossover=}).
The energies are measured in the units of $(\gamma_0^2\Delta/2)^{1/3}$.
}
\end{figure}

When the right-hand side of Eq.~(\ref{sqrtSchr=}) is neglected,
by a substitution $y=-(4\alpha)^{-2/3}z^2$ it is reduced to the
Airy equation, so the linearly independent solutions are the
derivatives of the Airy functions:
\begin{equation}\label{tpsi=AiBi}
\tilde\psi(y)
=C_\mathrm{A}\Ai'\!\left((4\alpha)^{1/3}\sqrt{-y}\right)
+C_\mathrm{B}\Bi'\!\left((4\alpha)^{1/3}\sqrt{-y}\right),
\end{equation}
valid at $|y|\ll\alpha^2$. The coefficients
$C_\mathrm{A},C_\mathrm{B}$ should
be determined by matching the WKB solutions, as described above,
at $\alpha^{2/3}\ll|y|\ll\alpha^2$.
At $y<0$, expanding
$\sigma(y)=\pi\alpha^2/4-(4/3)\sqrt\alpha(-y)^{3/4}+O((-y)^{5/4}$
and using the asymptotic expression
$\Bi'(z)\approx{z}^{1/4}e^{-(2/3)z^{3/2}}/\sqrt{\pi}$ for real $z>0$,
we obtain $C_\mathrm{B}=\sqrt\pi(2\alpha^2)^{-1/6}e^{-\pi\alpha^2/4}$.
At $y>0$, the exponentially decaying linear combination of $\Ai'(-is)$
and $\Bi'(-is)$ with real $s>0$ is obtained by taking
$C_\mathrm{A}=iC_\mathrm{B}$. 

Now, to find the exponentially small difference between
$|C_+|$ and $|C_-|$ in Eq.~(\ref{tpsiWKB=}) it is sufficient
to evaluate the current.
On the one hand, the current carried by solution~(\ref{tpsiWKB=})
at $E<E_1$ is given by $J=2\sqrt{wh}(|C_+|^2-|C_-|^2)$. On the other,
from Eq.~(\ref{continuity=}),
\begin{equation}
J=-2\sqrt{wh}
\int\limits_0^\infty\frac\alpha{\sqrt{y}}\,|\tilde\psi(y)|^2\,dy.
\end{equation}
For $\tilde\psi(y)$, it is sufficient to use
expression~(\ref{tpsi=AiBi}) as the integral converges at
$y\sim\alpha^{-2/3}\ll{1}$. This gives the leading exponential in
$1-p_\infty$:
\begin{eqnarray}
&&1-p_\infty\approx\pi{e}^{-\pi\alpha^2/2}
\int\limits_0^\infty\left|i\Ai'(-is)+\Bi'(-is)\right|^2ds=\nonumber\\
&&\hspace*{1.5cm}{}={e}^{-\pi\alpha^2/2},
\end{eqnarray}
which is Eq.~(\ref{padiabatic=}). The integral is calculated by parts:
\begin{eqnarray}
&&\int\limits_0^\infty\left|i\Ai'(-is)+\Bi'(-is)\right|^2ds=\nonumber\\
&&=\frac{1-i/\sqrt{3}}\pi
+\int\limits_0^\infty\left|\Ai(is)+i\Bi(is)\right|^2is\,ds.
\end{eqnarray}
While the first line is purely real,
the last integral is purely imaginary,
so the first line must be equal to $1/\pi$.

\section*{Numerical solution of the dynamical problem}
\label{app:numerical}

\begin{figure}[t]
\includegraphics[width=8.5cm]{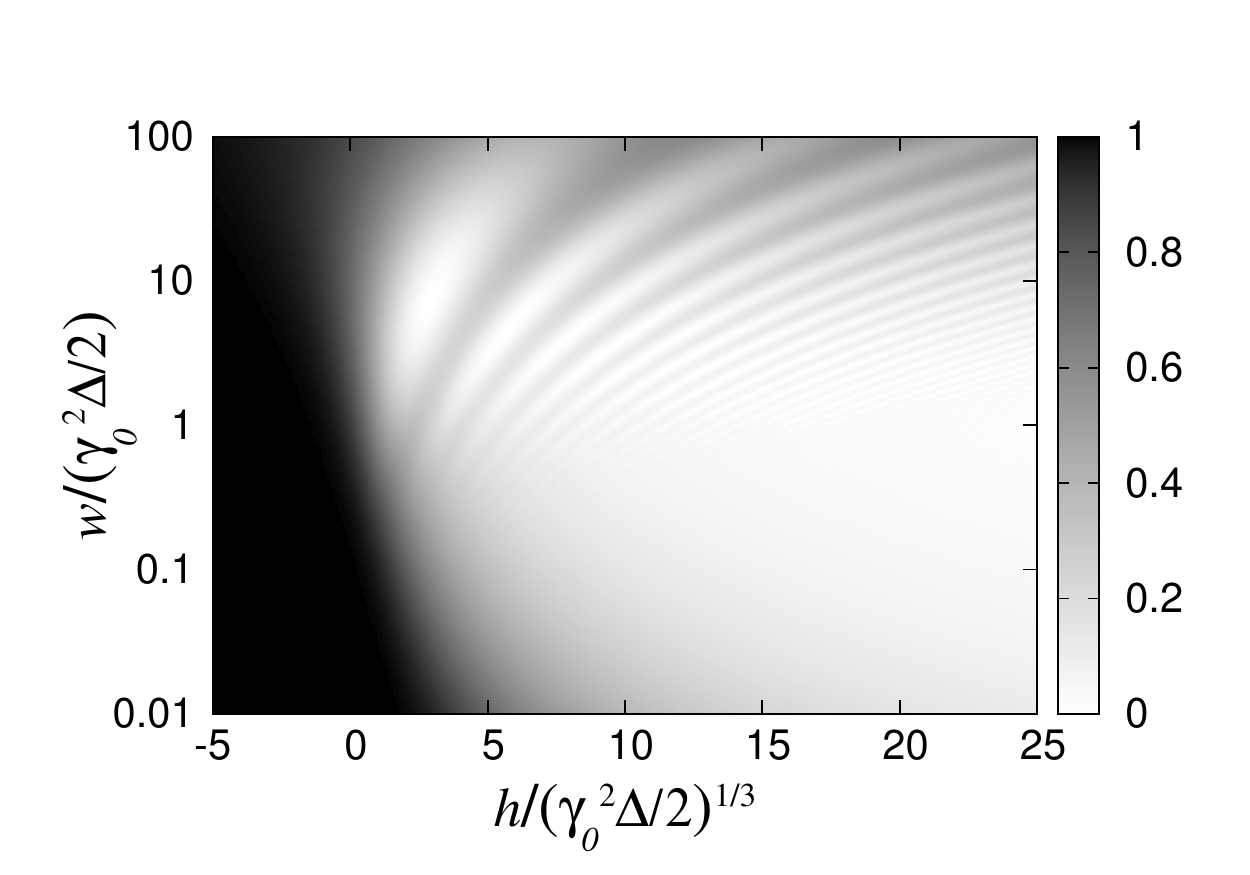}
\caption{\label{fig:parabolic2D} 
The numerical results for $p_\infty$ in the whole $(h,w)$ plane.
The black and white color correspond to $p_\infty\to{1}$ and
$p_\infty\to{0}$, respectively.
The energies are measured in the units of $(\gamma_0^2\Delta/2)^{1/3}$.
}
\end{figure}

To solve the dynamical problem numerically, it is more convenient
to return to the original problem~(\ref{ddtpsi=}),~(\ref{ddtphi=}),
rather than to integrate Eq.~(\ref{dpsidt=}) with the long-range
memory kernel. Equations~ (\ref{ddtpsi=}) and ~(\ref{ddtphi=}) can be
equivalently rewritten in the coordinate representation, which
is implemented numerically as a tight-binding model:
\begin{eqnarray}
i\,\frac{\partial\psi_\md}{\partial{t}}&=&E_\md(t)\,\psi_\md+V\phi_0,\\
i\,\frac{\partial\phi_n}{\partial{t}}&=&\delta_{n0}V\psi_\md
+J(2\phi_n-\phi_{n-1}-\phi_{n+1}).
\end{eqnarray}
To determine the coefficients, it is convenient to consider the
self-energy for the tight-binding model,
\begin{equation}
\Sigma^\mathrm{TB}(E)=\int\limits_{-\pi}^\pi
\frac{V^2}{E-2J(1-\cos{k})}\,\frac{dk}{2\pi}
=\frac{V^2\sign(E-2J)}{\sqrt{E(E-4J)}},
\end{equation}
and to match the coefficient at $1/\sqrt{E}$,
which gives
\begin{equation}
\frac{V^2}{\sqrt{4J}}=\sqrt{\frac{\gamma_0^2\Delta}2}.
\end{equation}
This still leaves a freedom of simultaneous rescaling of
$V$ and $J$ which keeps $V^2/\sqrt{J}$ constant. The value
of $J$ should be chosen so that the level energy is always
in the parabolic part of the spectrum,
$J\gg\max\{E_\md,(\gamma_0^2\Delta)^{1/3}\}$.

It is convenient to impose periodic boundary conditions,
$\phi_{n=N-1}=\phi_{n=-(N-1)}$,
so that the chain has $2N-2$ sites, and to notice that the
$(N-2)$-dimensional odd subspace $\phi_n=-\phi_{-n}$ decouples
from the discrete level.
Thus, one can work with the $N$-dimensional even subspace,
for which the equations become
\begin{eqnarray}
&&i\,\frac{\partial\phi_0}{\partial{t}}=V\psi_\md
+2J(\phi_0-\phi_1),\\
&&i\,\frac{\partial\phi_n}{\partial{t}}=
J(2\phi_n-\phi_{n-1}-\phi_{n+1}),\quad 0<n<N-1,\qquad\\
&&i\,\frac{\partial\phi_{N-1}}{\partial{t}}=2J(\phi_{N-1}-\phi_{N-2}).
\end{eqnarray}
The eigenstates of the unperturbed problem are
\begin{equation}
a_{kn}\propto\cos\frac{\pi{k}n}{N-1},\quad
E_k=2J\left(1-\cos\frac{\pi{k}}{N-1}\right),
\end{equation}
with $k=0,1,\ldots,N-1$.
The length of the chain should be sufficiently large,
so that
\begin{equation}
E_{k=1}-E_{k=0}\approx{J}\frac{\pi^2}{N^2}
\ll|E_*|\approx\frac{\gamma_0^2\Delta}{2E_\md^2}.
\end{equation}

The resulting system of ordinary differential equations is
integrated using the Bulirsch-Stoer method with polynomial
extrapolation.
The results of the numerical integration for $E_\md(t)=h-wt^2$
are shown in Figs.~\ref{fig:curves} and~\ref{fig:parabolic2D},
where $(\gamma_0^2\Delta/2)^{1/3}$ is used as the natural unit
of energy. From Fig.~\ref{fig:curves} one can see that except
the region $h^3\sim{w}\sim\gamma_0^2\Delta$, the numerical
result is well captured by at least one of the three analytical
expressions, Eqs.~(\ref{pMarkovian=}),~(\ref{padiabatic=}),
and~(\ref{pcrossover=}). Remarkably, at $w\gg\gamma_0^2\Delta$,
the Stueckelberg oscillations are reproduced both by the Markovian
Eq.~(\ref{pMarkovian=}) and by the crossover Eq.~(\ref{pcrossover=}).

\end{document}